\documentclass[preprint]{ptephy_v1}

\preprintnumber{KYUSHU-HET-169}

\usepackage{amssymb}
\usepackage{amsmath}
\usepackage{mathtools}
\DeclareMathOperator{\tr}{tr}
\numberwithin{equation}{section}
\usepackage{graphicx}
\usepackage{hyperref}

\begin{document}

\title{Lorentz symmetry violation in the fermion number anomaly with the chiral
overlap operator}

\author{Hiroki Makino}
\author{Okuto Morikawa}
\affil{
Department of Physics, Kyushu University,
744 Motooka, Nishi-ku, Fukuoka, 819-0395, Japan
}

\begin{abstract}
Recently, Grabowska and Kaplan proposed a four-dimensional lattice formulation
of chiral gauge theories on the basis of a chiral overlap operator. We compute
the classical continuum limit of the fermion number anomaly in this
formulation. Unexpectedly, we find that the continuum limit contains a term which is
not Lorentz invariant. The term is, however, proportional to the gauge anomaly
coefficient, and thus the fermion number anomaly in this lattice formulation
automatically restores the Lorentz-invariant form when and only when the
anomaly cancellation condition is met.
\end{abstract}
\subjectindex{B01, B05, B31}
\maketitle

\section{Introduction}\label{sec:intro}
It is important to give a non-perturbative definition of chiral gauge theories.
Recently, Grabowska and Kaplan constructed a five-dimensional domain-wall
lattice formulation of chiral gauge theories~\cite{Grabowska:2015qpk}.\footnote{
For a six-dimensional domain-wall formulation related to their formalism,
see~Ref.~\cite{Fukaya:2016ofi}} More
recently, they proposed a four-dimensional lattice formulation on the basis of
the so-called chiral overlap operator which is derived from the above
domain-wall formulation~\cite{Grabowska:2016,Kaplan:2016}. Their
four-dimensional formulation contains left- and right-handed fermions and, in the
tree-level approximation, the left-handed component couples only to the original
gauge field and the right-handed component couples only to a gauge field
evolved by the gradient
flow~\cite{Narayanan:2006rf,Luscher:2009eq,Luscher:2010iy,Luscher:2011bx} for
infinite time. The right-handed Weyl fermion is called the fluffy mirror
fermion or fluff. Okumura and~Suzuki~\cite{Okumura:2016dsr} argued that the
fermion number anomaly in this formulation possibly has phenomenological
implications for the strong CP problem, baryogenesis, and the dark matter
problem. They also conjectured the form of the classical continuum limit of the
fermion number anomaly, but the explicit calculation was not carried out
in~Ref.~\cite{Okumura:2016dsr}.

In the present paper, we complete the calculation of the classical continuum
limit of the fermion number anomaly in the formulation
of~Refs.~\cite{Grabowska:2016,Kaplan:2016}; the correct expression turns out
to be more complicated than the simple expression conjectured
in~Ref.~\cite{Okumura:2016dsr}. Rather unexpectedly, we find that the anomaly
contains a term which is not Lorentz invariant. The term is proportional to the
gauge anomaly coefficient and thus the fermion number anomaly in this lattice
formulation automatically restores the Lorentz-invariant form when and only
when the anomaly cancellation condition is met. The physical meaning of
this finding is not immediately obvious; however, remembering that the fermion
number anomaly is a very basic property of chiral gauge theories and any
sensible formulation of chiral gauge theories must fail when the anomaly
cancellation condition is not met, our finding appears interesting and quite
suggestive.

\section{Basic formulation}\label{sec:formulation}
In the formulation of~Ref.~\cite{Grabowska:2016,Kaplan:2016}, there are two
gauge fields, $A$ and~$A_\star$. $A$ couples to the physical left-handed
fermion while $A_\star$ is given from~$A$ by the gradient flow for infinite
flow time and couples to the would-be invisible right-handed fermion, the
fluffy mirror fermion. This formulation manifestly preserves the gauge
invariance. If we regard the gauge fields as non-dynamical external fields, the
partition function is given by
\begin{align}
 \int \mathcal{D}\psi \mathcal{D}\Bar\psi
 \exp\left[- a^4 \sum_x \Bar\psi(x) \mathcal{D}_\chi \psi(x)\right],
\end{align}
where $a$ is the lattice spacing and $\mathcal{D}_\chi$ denotes the chiral
overlap operator,
\begin{align}
 a \mathcal{D}_\chi
 = 1 + \gamma_5 \left[1 -
 (1 - \epsilon_\star) \frac{1}{\epsilon\epsilon_\star + 1} (1 - \epsilon)\right].
\end{align}
Here we have used the sign functions
\begin{align}
 \epsilon = \frac{H_w[A]}{\sqrt{H_w[A]^2}},\quad
 \epsilon_\star = \frac{H_w[A_\star]}{\sqrt{H_w[A_\star]^2}},
\end{align}
of the Hermitian Wilson Dirac operator
\begin{align}
 H_w
 = \gamma_5 \left[\frac{1}{2} \gamma_\mu (\nabla_\mu + \nabla_\mu^*)
 - \frac{1}{2} a \nabla_\mu \nabla_\mu^* - m\right],
 \label{eq:def-H_w}
\end{align}
where $\nabla_\mu$ is the forward gauge-covariant lattice derivative and
$\nabla_\mu^*$ is the backward one,
\begin{align}
 \nabla_\mu[A] f(x)
 &= \frac{1}{a}\left[U(x, \mu) f(x + a\Hat\mu) - f(x)\right]
 \label{eq:nabla}\\
 &= \left[D_\mu + \frac{a}{2} D_\mu D_\mu + \frac{a^2}{6} D_\mu D_\mu D_\mu + \mathcal{O}\left(a^3\right)\right] f(x),
 \label{eq:nabla_cont}\\
 \nabla_\mu^*[A] f(x)
 &= \frac{1}{a}\left[f(x) - U(x - a\Hat\mu, \mu)^\dagger f(x - a\Hat\mu)\right]
 \label{eq:nabla*}\\
 &= \left[D_\mu - \frac{a}{2} D_\mu D_\mu + \frac{a^2}{6} D_\mu D_\mu D_\mu + \mathcal{O}\left(a^3\right)\right] f(x).
 \label{eq:nabla*_cont}
\end{align}
In Eqs.~\eqref{eq:nabla} and~\eqref{eq:nabla*}, the link variable is given by
\begin{align}
 U(x, \mu)[A] = P \exp\left[a \int_0^1 dt A_\mu(x + t a \Hat\mu)\right],
\end{align}
where $P$ denotes the path-ordered product
and $\Hat\mu$ is the unit vector in the direction of~$\mu$;
in~Eqs.~\eqref{eq:nabla_cont} and~\eqref{eq:nabla*_cont},
$D_\mu = \partial_\mu + A_\mu$.
For $\nabla_\mu[A_\star]$ and $\nabla_\mu^*[A_\star]$,
$D_\mu$ is replaced by $D_{\star\mu} = \partial_\mu + A_{\star\mu}$.
The sign functions satisfy
\begin{align}
 \epsilon^2 = \epsilon_\star^2 = 1,\quad
 \left[1 - (1 - \epsilon_\star) \frac{1}{\epsilon\epsilon_\star + 1} (1 - \epsilon) \right]^2
 = 1
\label{eq:(2.5)}
\end{align}
and, as a consequence, the Ginsparg--Wilson relation~\cite{Ginsparg:1981bj}
\begin{align}
 \gamma_5 \mathcal{D}_\chi + \mathcal{D}_\chi \gamma_5
 = a\mathcal{D}_\chi \gamma_5 \mathcal{D}_\chi
\end{align}
holds. It is then natural to introduce a modified
$\gamma_5$~\cite{Okumura:2016dsr,Luscher:1998pqa,Niedermayer:1998bi}
\begin{align}
 \Hat\gamma_5 \equiv \gamma_5 (1 - a\mathcal{D}_\chi)
 \label{eq:Hatgamma_5}
\end{align}
which satisfies
\begin{align}
 \left(\Hat\gamma_5\right)^2 =1,
 \quad
 \mathcal{D}_\chi \Hat\gamma_5 = - \gamma_5 \mathcal{D}_\chi.
 \label{eq:Hatgamma_5_pro}
\end{align}
Note that $\Hat\gamma_5$ is not Hermitian in this formulation.
Using modified chiral projection operators
\begin{align}
 \Hat{P}_\pm \equiv \frac{1}{2} \left(1 \pm \Hat\gamma_5\right),
\end{align}
the chiral components of the fermion can be defined as
\begin{align}
 \Hat{P}_- \psi_L(x) &= \psi_L(x),&
 \Bar\psi_L(x) P_+ &= \Bar\psi_L(x),
\label{eq:(2.10)}
\\
 \Hat{P}_+ \psi_R(x) &= \psi_R(x),&
 \Bar\psi_R(x) P_- &= \Bar\psi_R(x).
\end{align}
Owing to the second relation in~Eq.~\eqref{eq:Hatgamma_5_pro}, the action is
decomposed into left- and right-handed components as
\begin{align}
 a^4 \sum_x \Bar\psi(x) \mathcal{D}_\chi \psi(x)
 &= a^4 \sum_x \left[\Bar\psi_L(x) \mathcal{D}_\chi \psi_L(x)
 + \Bar\psi_R(x) \mathcal{D}_\chi \psi_R(x)\right].
\end{align}

\section{The classical continuum limit of the fermion number anomaly}
\label{sec:fermion-number-anomaly}
The fermion number anomaly on the lattice associated with the left-handed
fermion in~Eq.~\eqref{eq:(2.10)} is given by~\cite{Okumura:2016dsr}
\begin{align}
 \mathcal{A}_L^{(a)}(x)
 \equiv \langle \partial_\mu j_{L\mu}(x) \rangle
 = \tr\left[\Hat{P}_-(x, x) - P_+\frac{1}{a^4}\delta_{x, x}\right]
 = - \frac{1}{2} \tr\Hat\gamma_5(x,x),
\end{align}
where $\tr$ stands for the trace over the spinor and gauge indices and we have
used $\tr\gamma_5=0$ to obtain the last expression.\footnote{We use the notation
$O(x, y) \equiv a^{-4} O_{x, y}$ for any matrix $O$.}
In what follows, we compute the classical continuum limit,
$\mathcal{A}_L\equiv\lim_{a\to0}\mathcal{A}_L^{(a)}$, for a smooth gauge field
configuration.

Let us first determine a general form of~$\mathcal{A}_L$, by assuming that it
is Lorentz invariant.
The following argument is helpful to simplify the explicit tedious calculation of~$\mathcal{A}_L$.
First, using Eq.~\eqref{eq:(2.5)}, we decompose
$\mathcal{A}_L^{(a)}$ into the parity-odd part~$\mathcal{A}_L^{(a)\text{odd}}$ and
the parity-even part~$\mathcal{A}_L^{(a)\text{even}}$ as
$\mathcal{A}_L^{(a)}=\mathcal{A}_L^{(a)\text{odd}}+\mathcal{A}_L^{(a)\text{even}}$,
where
\begin{align}
 \mathcal{A}_L^{(a)\text{odd}}(x)
 &= \frac{1}{2} \tr \frac{2}{\epsilon + \epsilon_\star}(x, x),&
 \mathcal{A}_L^{(a)\text{even}}(x)
 &= \frac{1}{2} \tr (\epsilon - \epsilon_\star)
 \frac{1}{\epsilon + \epsilon_\star}(x, x).
\end{align}
Then it is obvious that, under the exchange of $A$ and~$A_\star$,
\begin{align}
 \mathcal{A}_L^{(a)\text{odd}}[A_\star, A]
 &= +\mathcal{A}_L^{(a)\text{odd}}[A, A_\star],&
 \mathcal{A}_L^{(a)\text{even}}[A_\star, A]
 &= -\mathcal{A}_L^{(a)\text{even}}[A, A_\star].
\end{align}
As the second property, we note when $A_\star=A$,
\begin{align}
 \mathcal{A}_L^{(a)}(x)[A, A]
 = \frac{1}{2} \tr\epsilon(x,x).
\end{align}
Finally, the integral of~$\mathcal{A}_L^{(a)}(x)$ over four-dimensional
spacetime is given by~\cite{Okumura:2016dsr}
\begin{align}
 a^4 \sum_x \mathcal{A}_L^{(a)}(x)
 &= \frac{1}{2} a^4 \sum_x \tr \epsilon(x,x),
\end{align}
and~\cite{Kikukawa:1998pd,Fujikawa:1998if,Adams:1998eg,Suzuki:1998yz}
\begin{align}
 \frac{1}{2}\tr\epsilon(x,x)
 &\stackrel{a\to0}{\longrightarrow}
 -\frac{1}{32\pi^2} \epsilon_{\mu\nu\rho\sigma}
 \tr\left[F_{\mu\nu}F_{\rho\sigma}\right],
\end{align}
for~$0 < ma <2$.

Now, for convenience, we introduce
\begin{align}
 C_\mu(x) \equiv A_{\star\mu}(x) - A_\mu(x),
\end{align}
which transforms as the adjoint representation under the gauge transformation
on $A$ and~$A_\star$.\footnote{In the effective action,
there could be gauge-invariant relevant operators in terms of $C_\mu$
such as the mass term $(1/a)^2 \tr C_\mu C_\mu$.
These terms would require fine-tuning toward the correct continuum limit.
We would like to thank the referee for a comment on this point.}
We note that $\mathcal{A}_L$ is
a dimension~$4$ gauge-invariant local polynomial of~$A$ and~$A_\star$.
Then, by examining the above properties,
we find that the most general form of~$\mathcal{A}_L$ is given
by
\begin{align}
 \mathcal{A}_L
 &= \Bar{\mathcal{A}}_L
 + d_1 \partial_\mu \tr\left[C_\mu C_\nu C_\nu\right]
 + \frac{1}{2} d_2\, \partial_\mu \tr\left[C_\nu \left\{F_{\mu\nu} + F_{\star\mu\nu}\right\}\right]
 \notag\\&\qquad
+ \text{(Lorentz symmetry violating part)},
 \label{eq:A_L-GL}
\end{align}
where
\begin{align}
 \Bar{\mathcal{A}}_L
 \equiv - \frac{1}{64\pi^2} \epsilon_{\mu\nu\rho\sigma}
 \left\{ \tr\left[F_{\mu\nu}F_{\rho\sigma}
 + F_{\star\mu\nu}F_{\star\rho\sigma}\right]
 - b\, \partial_\mu \tr\left[C_\nu \mathcal{D}_\rho C_\sigma 
 + C_\nu \mathcal{D}_{\star\rho} C_\sigma\right]\right\}
 \label{eq:A_L^chi}
\end{align}
with $\mathcal{D}_\rho = \partial_\rho + [A_\rho, \cdot]$
and~$\mathcal{D}_{\star\rho} = \partial_\rho + [A_{\star\rho}, \cdot]$. The
coefficients $d_1$, $d_2$, and~$b$ cannot be determined from the above argument
alone. In the first line of~Eq.~\eqref{eq:A_L-GL},
$\Bar{\mathcal{A}}_L$~\eqref{eq:A_L^chi} arises from the parity-odd part and
the other terms from the parity-even part. The second
term~$\partial_\mu \tr[C_\mu C_\nu C_\nu]$ is proportional to the gauge anomaly
coefficient and thus it vanishes for anomaly-free cases. The following
explicit calculation shows that~$d_2=0$ in the third term. As we will show
below, there actually exists a Lorentz symmetry violating term
in~$\mathcal{A}_L$. Note that, generally speaking, the restoration of the
Lorentz symmetry is not automatic with the lattice regularization.

Let us describe how the explicit calculation of~$\mathcal{A}_L$
proceeds.\footnote{We used the \texttt{Mathematica} package \texttt{NCAlgebra}
for this calculation.} We first note that
\begin{align}
 \tr\Hat\gamma_5(x, x)
 = \sum_y \tr\Hat\gamma_5(x, y) \delta_{y,x}.
\end{align}
In this expression, we use
\begin{align}
 \delta_{y,x}
 = \int_{-\pi}^\pi \frac{d^4p}{(2\pi)^4} e^{ip(y - x) / a}
 \equiv \int_p e^{ip(y - x) / a}.
\end{align}
From Eq.~\eqref{eq:def-H_w}, we have
\begin{align}
 &\sum_y a H_w(x, y)[A] e^{ipy/a} f(y)\notag\\
 &= e^{ipx/a} \gamma_5
 \sum_y \left[i \sum_\mu \gamma_\mu (s_\mu - iaQ_\mu)
 - \sum_\mu (c_\mu - 1) - aR - ma\right]\!\!(x, y)\, f(y),
\label{eq:(3.11)}
\end{align}
where
\begin{align}
 s_\mu &\equiv \sin p_\mu, &
 c_\mu &\equiv \cos p_\mu, \\
 Q_\mu &\equiv \frac{1}{2}\left(e^{ip_\mu}\nabla_\mu + e^{-ip_\mu}\nabla_\mu^*\right), &
 R &\equiv \frac{1}{2}\sum_\mu \left(e^{ip_\mu}\nabla_\mu - e^{-ip_\mu}\nabla_\mu^*\right).
\label{eq:Q-R}
\end{align}
$Q_{\star\mu}$ and~$R_\star$ are defined similarly. Thus, $\mathcal{A}_L^{(a)}$ can be
written in terms of operators $Q_\mu$, $R$, $Q_{\star\mu}$, and~$R_\star$. Next,
we expand $\mathcal{A}_L^{(a)}$ into the power series
of the lattice spacing $a$ up to~$O(a^0)$,
noting that $ma \sim \mathcal{O}\left(a^0\right)$.
For this, we need the following
expansions:
\begin{align}
 Q_\mu
 &= c_\mu D_\mu
 + \frac{a}{2} i s_\mu D_\mu D_\mu
 + \frac{a^2}{6} c_\mu D_\mu D_\mu D_\mu
 + \mathcal{O}\left(a^3\right),\\
 R
 &= \sum_\mu \left(
 is_\mu D_\mu
 + \frac{a}{2} c_\mu D_\mu D_\mu
 + \frac{a^2}{6} is_\mu D_\mu D_\mu D_\mu
 + \mathcal{O}\left(a^3\right)\right).
\end{align}

Carrying out the explicit calculation, we
find that only four covariant derivatives with the same spacetime indices appear in
the Lorentz symmetry violating terms. Therefore, taking into account the above
properties of the fermion number anomaly, the Lorentz symmetry violating part
in the general form~\eqref{eq:A_L-GL} must be
\begin{align}
 \text{(Lorentz symmetry violating part)}
 = d'_1 \partial_\mu \tr\left[C_\mu C_\mu C_\mu\right],
 \label{eq:A_L-LV}
\end{align}
with a to-be-determined coefficient~$d'_1$.

\begin{figure}
 \centering
 \begin{minipage}{0.45\columnwidth}
  \centering
  \includegraphics[width=\columnwidth]{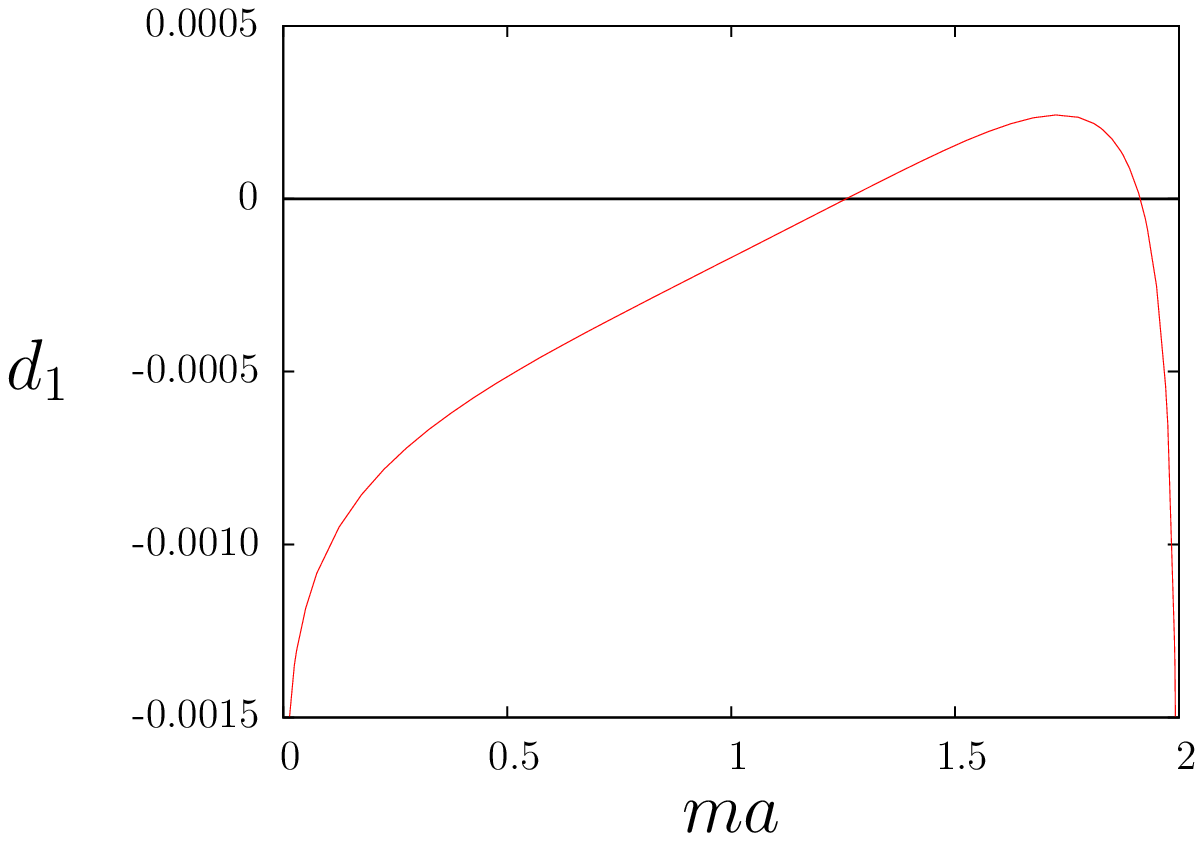}
  \caption{$d_1(ma)$}
  \label{fig:LSd5}
 \end{minipage}
 \hspace{1em}
 \begin{minipage}{0.45\columnwidth}
 \centering
 \includegraphics[width=\columnwidth]{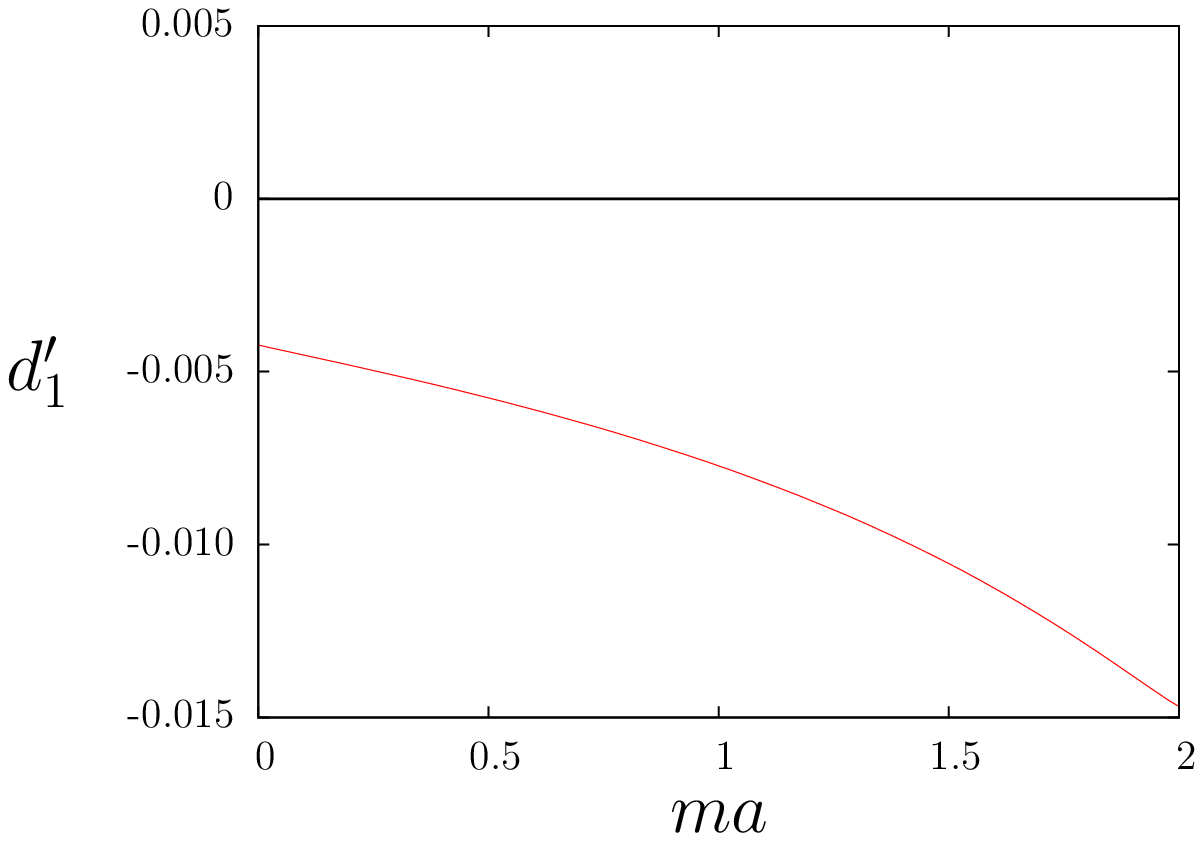}
 \caption{$d'_1(ma)$}
 \label{fig:LVd5}
 \end{minipage}
\end{figure}

Then, the explicit tedious expansion of~$\mathcal{A}_L$ can be exactly
combined into the form of~Eqs.~\eqref{eq:A_L-GL} and~\eqref{eq:A_L-LV}.
After this calculation, we finally obtain
\begin{align}
 \mathcal{A}_L(x)
 &=
 \Bar{\mathcal{A}}_L(x)
 + d_1 \partial_\mu \tr\left[C_\mu C_\nu C_\nu\right]
 + d'_1 \partial_\mu \tr\left[C_\mu C_\mu C_\mu\right],
 \label{eq:A_L-Result}
\end{align}
where $\Bar{\mathcal{A}}_L(x)$ is given by~Eq.~\eqref{eq:A_L^chi}
with the correct overall factor,\footnote{The momentum integral in this factor
is known in the context of the axial anomaly with the usual overlap operator.
See, for example,~Ref.~\cite{Fujiwara:2002xh} and the references cited therein.}
and the coefficients are
\begin{align}
 b &= \frac{2}{3},\label{eq:b}\\
 d_1(ma)
 &= 
 \frac{1}{128} \int_p \frac{1}{t} c_\rho c_\sigma
 + \frac{1}{8}\int_p \frac{1}{t^2}
 - \frac{1}{32} \int_p \frac{1}{t^2} s_\rho^2 s_\rho^2\notag\\
 &\qquad
 + \frac{3}{16} \int_p \frac{1}{t^2} (c - c_\rho) c_\rho
 + \frac{3}{64} \int_p \frac{1}{t^2} (c - c_\rho) (c - c_\sigma) c_\rho c_\sigma,
 \label{eq:d_2}\\
 d'_1(ma)
 &= 
 - \frac{1}{12} \int_p \frac{1}{t}
 - \frac{1}{128} \int_p \frac{1}{t} c_\rho c_\sigma
 + \frac{1}{192} \int_p \frac{1}{t} (c - c_\rho) c_\rho\notag\\
 &\qquad
 + \frac{1}{32} \int_p \frac{1}{t^2} s_\rho^2 s_\sigma^2
 + \frac{3}{64} \int_p \frac{1}{t^2} (c - c_\rho)^2
 - \frac{3}{64} \int_p \frac{1}{t^2} (c - c_\rho) (c - c_\sigma) c_\rho c_\sigma,
 \label{eq:d'_2}
\end{align}
and $d_2=0$ as mentioned above, where
\begin{align}
 c &\equiv \sum_\mu (c_\mu - 1) + ma, \quad
 t \equiv \sum_\mu s_\mu^2 + c^2.
\end{align}
In the integrals in Eq.~\eqref{eq:d_2} and Eq.~\eqref{eq:d'_2}, indices $\rho$
and~$\sigma$ are arbitrary as long as they differ from each other. The
coefficients $d_1(ma)$ and~$d'_1(ma)$ as functions of~$ma$ are plotted
in~Figs.~\ref{fig:LSd5} and~\ref{fig:LVd5}. As already announced, the last term
in~Eq.~\eqref{eq:A_L-Result} is not Lorentz invariant. However, this term is
proportional to the gauge anomaly coefficient. Thus, in anomaly-free chiral
gauge theories, this Lorentz symmetry violating term [and the second term
of~Eq.~\eqref{eq:A_L-Result}] vanishes; only $\Bar{\mathcal{A}}_L$ provides
the fermion number anomaly.

\section{Conclusion}\label{sec:conclusion}
In the present paper, we computed the classical continuum limit of the
fermion number anomaly in the lattice formulation of chiral gauge theories by
Grabowska and Kaplan. The anomaly consists of two parts: One is the parity-odd
part being proportional to the epsilon tensor, $\Bar{\mathcal{A}}_L$. The
other is the parity-even part, which is proportional to the gauge anomaly
coefficient. The latter contains a Lorentz symmetry violating term. In
anomaly-free cases, only the former~$\Bar{\mathcal{A}}_L$ contributes, which is
Lorentz invariant. It appears quite interesting and suggestive that the Lorentz
symmetry and the gauge anomaly are linked in this way in this lattice
formulation on the basis of the fluffy mirror fermion.

\section*{Acknowledgment}
We are grateful to Hiroshi Suzuki for his support, helpful advice, and a careful
reading of the manuscript.

\section*{Note added}
In the present paper, we considered only the fermion number $U(1)$.
As is discussed in Ref.~\cite{Grabowska:2016bis},
however, the gauge anomaly cancellation condition does not necessarily imply
the vanishing of the $d'_1$ term of Eq.~\eqref{eq:A_L-Result} for more general $U(1)$ charges.

\end{document}